\documentclass[twocolumn,prd]{revtex4}
\usepackage{amsfonts,amsbsy,amssymb,amsmath,graphicx,float,subfigure,color} 
\usepackage[linktoc=all,colorlinks=true,linkcolor=blue,citecolor=blue]{hyperref}
\usepackage{mathrsfs}
\pdfoutput=1
\usepackage{graphicx}
\usepackage{breqn}    % to match \left and \right across line breaks (\\)
\newcommand{\abs}[1]{\left\lvert#1\right\rvert}

\newcommand{\Poincare}{Poincar\'e\ }
\DeclareMathOperator{\tr}{tr}
\newcommand{\hc}{h_{\mathrm{c}}}
\newcommand{\Tee}{\mathrm{T}}
\begin{document}

\title{Stability of Leapfrogging Vortex Pairs: A Semi-analytic Approach}

\author{Brandon M. Behring} 
\email{bmb29@njit.edu}
\affiliation{Department of Mathematical Sciences$,$ New Jersey Institute of Technology\\ University Heights Newark$,$ NJ 07102}

\author{Roy H. Goodman}
\email{goodman@njit.edu}
\affiliation{Department of Mathematical Sciences$,$ New Jersey Institute of Technology\\ University Heights Newark$,$ NJ 07102}

\begin{abstract}
We investigate the stability of a one-parameter family of periodic solutions of the four-vortex problem known as `leapfrogging' orbits. These solutions, which consist of two pairs of identical yet oppositely-signed vortices, were known to W.\ Gr\"obli (1877) and  A.\ E.\ H.\ Love (1883), and can be parameterized by a dimensionless parameter $\alpha$  related to the geometry of the initial configuration. Simulations by Acheson (2000) and numerical Floquet analysis by Toph{\o}j and Aref (2012) both indicate, to many digits, that the bifurcation occurs when $1/\alpha=\phi^2$, where $\phi$ is the golden ratio. This study aims to explain the origin of this remarkable value. Using a trick from the gravitational two-body problem, we change variables to render the Floquet problem in an explicit form that is more amenable to analysis. We then implement G. W. Hill's method of harmonic balance to high order using computer algebra to construct a rapidly-converging sequence of asymptotic approximations to the bifurcation value, confirming the value found earlier.
\end{abstract}
\maketitle
%\tableofcontents
 \section{Introduction}
Point-vortex motion arises in the study of concentrated vorticity in an ideal, incompressible fluid described by Euler's equations. The two-dimensional Euler equations of fluid mechanics, a partial differential equation (PDE) system, support a solution in which the vorticity is concentrated at a single point. Helmholtz derived a system of ordinary differential equations (ODEs) that describe the motion of a set of interacting vortices that behave as discrete particles, which approximates the fluid motion in the case that the vorticity is concentrated in very small regions~\cite{Helmholtz}. This system of equations has continued to provide interesting questions for over $150$ years.  For a thorough introduction and review see Refs.~\cite{newton2013n,Aref2007,Aref150}. 

Kirchhoff formulated these equations as a Hamiltonian system~\cite{Aref2007,kirchhoff1876}. This has allowed researchers to apply to this system a wide repertoire of methods that were developed in the study of the gravitational $N$-body problem. In this paper, we consider a configuration of vortices with vanishing total circulation, which has no analogue in the $N$-body problem. As such, many techniques developed for the gravitational problem do not  apply to the net-zero circulation case of the $N$-vortex problem. Because of this, this case of the $N$-vortex problem is relatively less studied, despite its physical importance and mathematical richness,  Ref.~\cite{Aref1992,ArefEckhardt,Aref1989}.

Bose-Einstein condensates (BEC), a quantum state of matter that exists at ultra-low temperatures, have provided an experimental testbed in which point vortices can be studied in the laboratory. These were first observed experimentally in Ref.~\cite{anderson1995} in 1995, work that led to the 2001 Nobel Prize in Physics for Cornell and Wieman, along with Ketterle. The same group experimentally demonstrated concentrated vortices in BECs~\cite{Matthews1999}. This has led in the last 20 years to a new flowering of interest in point vortices. In this experimental system, the BEC is confined using a strong magnetic field that introduces additional terms into the equations of motion. Ref.~\cite{Navarro:2013hb}, for example, shows nicely how experiment and mathematical theory have been used together to explore these nonlinear phenomena.

\begin{figure*}[thbp] %  figure* placement: here, top, bottom, or page
 
 \includegraphics[width=.7\textwidth]{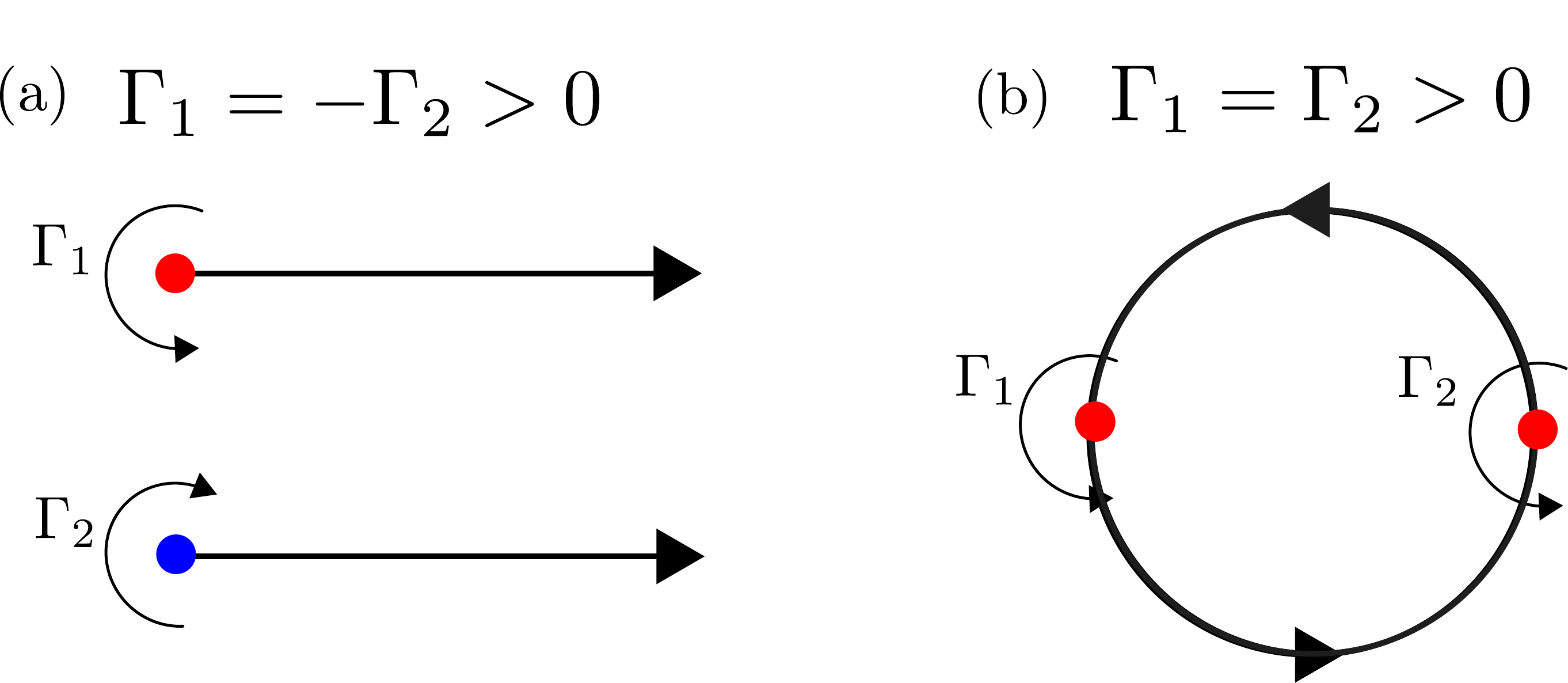} 
 \caption{(a) Opposite-signed vortices move in parallel along straight lines. (b) Like-signed vortices move along a circular path.}
\label{fig:two_vortex}
\end{figure*}

The leapfrogging solution to the point-vortex system of equations is built from simple components. As is well-known, two vortices of equal and of opposite-signed vorticity  move in parallel at a uniform speed with their common velocity inversely proportional to the distance between them. Two vortices of equal and like-signed vorticity, by contrast, trace a circular path with a constant rotation rate proportional to the inverse square of the distance between them, see Fig.~\ref{fig:two_vortex}.

\begin{figure}[ht]
   \includegraphics[width=.5\textwidth]{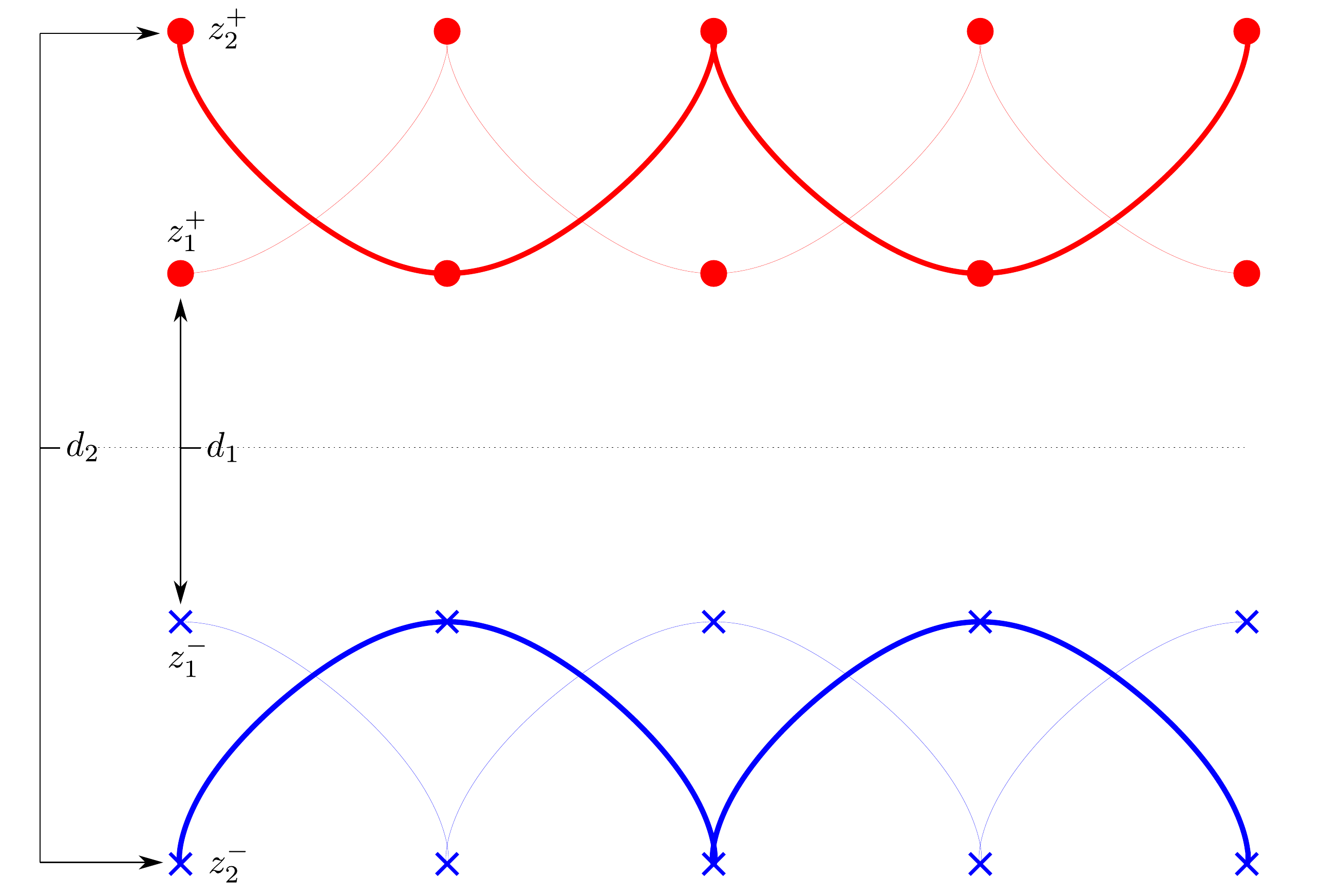}
\caption{Motion in physical ($z$) space. Average motion is from left to right. Markers are given every half-period.}
\label{fig:configuration}
    \end{figure}

Now consider a system of four vortices of equal strength, arranged collinearly and symmetrically at $t=0$, with vortices of strength positive one at $z_1^{+}$ and $z_2^{+}$ and vortices of strength negative one at $z_1^{-}$ and $z_2^{-}$; see Fig.~\ref{fig:configuration}. Throughout this paper we represent particle positions by points in the complex $z$ plane. Let the `breadths' of the pairs  denote the distances $d_1=|z_1^{+}-z_1^{-}|$  and $d_2=|z_2^{+}-z_2^{-}|>d_1$ at $t=0$. This symmetric collinear state depends, after a scaling, on only one dimensionless parameter, the ratio of the breadths of the pairs, $\alpha=d_1/d_2$.

This configuration provides the initial condition for a remarkable family of relative periodic orbits known as `leapfrogging orbits', described first by Gr\"{o}bli in 1877~\cite{grobli} and independently by Love (1883)~\cite{Love}. It can be considered as a simple two-dimensional model of the phenomenon of two smoke rings passing through each other periodically, first discussed by Helmholtz in 1858~\cite{Helmholtz,Borisov}. Recall that a relative periodic orbit is defined as an orbit that is periodic modulo a group orbit of a symmetry of the system, in this case translation.

With reference to Fig.~\ref{fig:configuration}, the two vortices $z_1^+$ and $z_1^-$ starting closer to the center of symmetry initially have larger rightward velocity than the outer pair, $z_2^+$ and $z_2^-$. As the `inner pair' propagates, the distance between them increases, causing them to slow down. Simultaneously, the distance between the `outer pair' decreases, causing them to speed up. After half a period, the identities of the inner and outer pairs are interchanged and the process repeats.  This relative periodic motion exists only for a finite range of breadth-ratios $\alpha_0 < \alpha < 1$ where $\alpha_0=3-2\sqrt{2}\approx 0.171573$.
\begin{figure*}[ht]
    \centering
\includegraphics[width=1.0\textwidth]{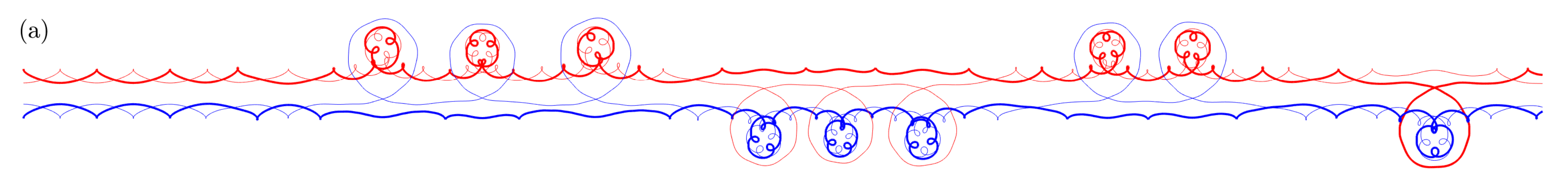}\\
\includegraphics[width=1.0\textwidth]{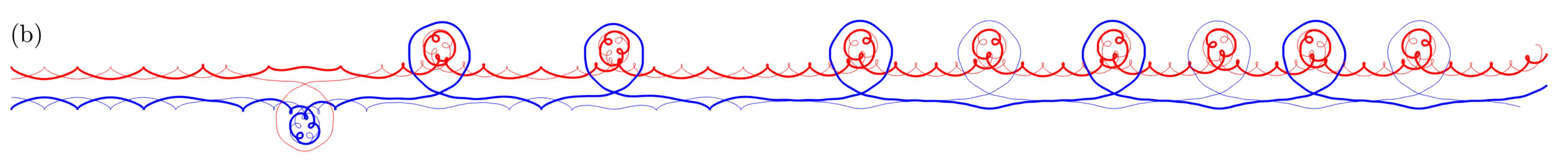}\\
\includegraphics[width=0.48\linewidth]{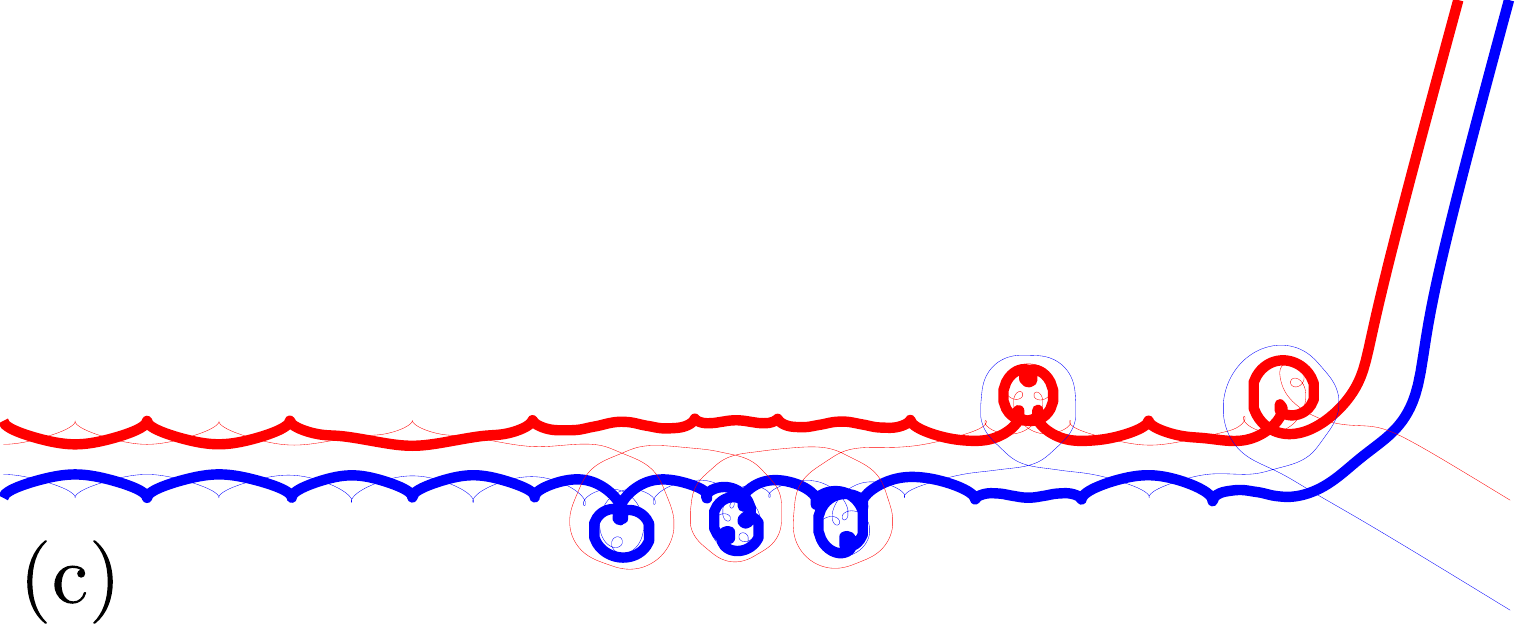}
\includegraphics[width=0.48\linewidth]{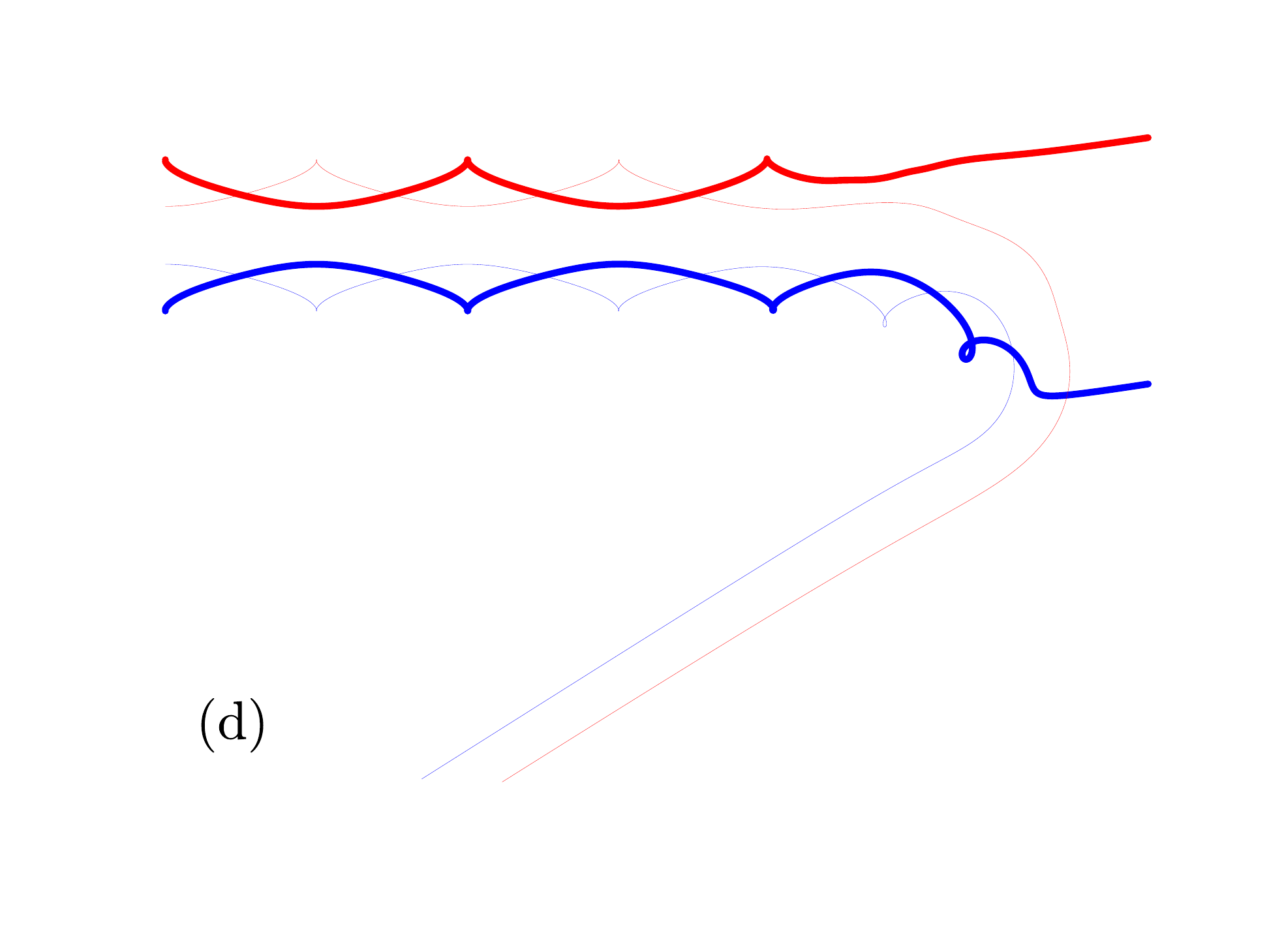}
\caption{Motion in physical ($z$) space. (a) This solution features several bouts of walkabout motion including  one extended period of three consecutive walkabout `dances'. (b) In this solution the last period of walkabout is braided as the two negative (blue) vortices take turns orbiting the tightly bound pair of positive (red) vortices.(c) A leapfrogging motion that transitions to walkabout motion before disintegrating. (d) A leapfrogging motion that disintegrates without a walkabout stage. }
\label{fig:walk-braid}
    \end{figure*}
As $\alpha$  approaches one, the distance separating the members of each pair of like-signed vortices becomes small compared to the distance between the two pairs. Each pair of like-signed vortices rotates quickly in a nearly circular orbit about its center of vorticity, similar to that of Fig.~\ref{fig:two_vortex}(b). The velocity field due to this pair is asymptotically close to that of a single vortex of twice the vorticity. Thus each pair of vortices moves with a velocity approximately given by such a velocity field and the two pairs move approximately along parallel lines in a motion resembling that depicted in Fig.~\ref{fig:two_vortex}(a).  As the parameter $\alpha$ is decreased, the coupling between the four vortices is stronger, and the motion can no longer be so neatly decoupled into two weakly-interacting pairs. This can lead to instability as the pairs approach each other and interact strongly enough to pull the pairs apart. 

Direct numerical simulations by Acheson  suggest that the leapfrogging solution is  stable  only for $\alpha> \alpha_2=\phi^{-2}=\frac{3-\sqrt{5} } {2} \approx 0.38$, where $\phi$ is the golden ratio~\cite{Acheson2000}.  Acheson observed that, after an initial period of exponential separation due to linear instability, the perturbed solutions could transition into one of two behaviors: a bounded orbit he called `walkabout'  and an unbounded orbit he called `disintegration.' In the walkabout orbit, two  like-signed vortices couple together and the motion resembles that of a three-vortex system. In disintegration,  four vortices separate into two pairs---each pair  consisting one negative and one positive vortex---that escape to infinity along two transverse rays, see Fig.~\ref{fig:walk-braid}. Acheson noted that  disintegration seemed only to occur for ratios  $\alpha< \alpha_1\approx 0.29$. It should be clearly noted that the  analysis used in this paper is able only to distinguish between linearly stable and unstable periodic orbits and tells us nothing about the mechanism of escape, which is the subject of a paper currently in preparation by the authors.

Toph{\o}j and Aref, having noticed similar behavior in the chaotic scattering of identical point vortices~\cite{Aref2008}, studied the stability problem further~\cite{ArefTophoj}. They examined  linearized perturbations about the periodic orbit, thereby reducing the stability question to a Floquet problem. They confirm Acheson's value of $\alpha_2$ via the numerical solution of this Floquet problem. However, their attempt at a more mathematical derivation of the fortuitous value of $\alpha_2$ depends on an \emph{ad hoc} argument based on `freezing' the time-dependent coefficients at their value at $t=0$, a method that has been known to sometimes produce incorrect results~\cite{Meiss, markus1960global}. In addition, they note from numerical simulations that there does not exist a value of $\alpha$ precisely separating walkabout from disintegration behavior. Rather, both can occur at the same value of $\alpha$  depending on the form of the perturbation.

More recently, Whitchurch et al.~\cite{Whitchurch} examined the system through the extensive use of numerically calculated  \Poincare\ surfaces of section. They observe that the bifurcation at $\alpha=\alpha_2$ is of Hamiltonian pitchfork type. They also identify a third type of breakup behavior in addition to walkabout and disintegration, which they call braiding, see Fig.~\ref{fig:walk-braid}(b). The existence of such a motion is implicit in the earlier three-vortex work of Rott~\cite{Rott} and the chaotic scattering work of Toph{\o}j and Aref~\cite{Aref2008}.

% WHAT WE'RE PLANNING TO DO
No satisfactory analytical explanation for the special value of the bifurcation $\alpha=\alpha_2$ exists in the literature, as all previous explanations have relied on numerical solution of the initial value problem. In the present work, we rewrite the Floquet system in a form that  allows for further analysis and use this to provide a semi-analytic argument for the bifurcation value using the method of harmonic balance  evaluate the Hill's determinant for the linearized perturbation equations.

% THIS PAPER IS ORGANIZED...

The remainder of the paper is organized as follows. 
In Sec.~\ref{sec:SETUP}, we review the equations of motion for the $N$-vortex problem and a canonical reduction of the phase space from  four degrees-of-freedom to two.  
In Sec.~\ref{sec:leap_and_linear}  we discuss the leapfrogging solution and summarize some of its properties. Further, we write down the linearized perturbation equations about the leapfrog orbit and discuss the relevant Floquet theory needed to understand its stability. The coefficients in these linearized perturbation equations had not been given in explicit form before now.
In Sec.~\ref{sec:ExplicitFloquet}, we introduce canonical polar coordinates and rewrite the stability equations explicitly in terms of the polar angle variable. We then expand this solution in terms of a small parameter and introduce a change of variables that further simplifies the later analysis. 
In Sec.~\ref{sec:HillHBM}, we first review Hill's method of harmonic balance. We then implement it to high order in a computer algebra system, thereby constructing a systematic and semi-analytic approximation to the bifurcation value. 
Lastly, in Sec.~\ref{sec:conclusion}, we summarize our work and discuss avenues for further study.

\section{Equations of motion}\label{sec:SETUP}

In this section, we will review the Hamiltonian framework for the $N$-vortex problem and introduce a reduction due to Aref and Eckhardt~\cite{ArefEckhardt} and apply it to the leapfrogging problem. We use complex coordinates to label the locations of the $N$ vortices located at  $z_j(t)=x_j+ i y_j$ and denote their (signed) vorticities by $\Gamma_j$. 

The locations of the vortices, given as coordinates in the complex plane, evolve as a Hamiltonian system with Hamiltonian function over the conjugate variables $z_i$ and $z_i^\ast$,
%\begin{equation}\label{NVORTEXHAM}
\[
\mathcal{H}(z, z^\ast)=-\frac{1}{4\pi} \sum_{1\le i<j \le N}  \Gamma_i \Gamma_j   \log{\left[\left(z_i-z_j\right)\left(z^\ast_i-z^\ast_j\right)\right]}.
\]
with Poisson brackets
\[
\{z_j,z_k\}=\{z^\ast_j,z^\ast_k\}=0 \quad \textrm{and} \quad \{z_j,z_k^\ast\}= \frac{2\delta_{jk}}{i \Gamma_k}.
\]

%\end{equation} 
When written in real coordinates, the conjugate variables are the $x$- and $y$-components of the motion. That is, the position space and the phase space coincide.  This gives rise to a system of \emph{first order} equations of motion.
%\begin{equation}\label{Ham}
\[
 \Gamma_j\dot{z}_j=-2i \frac{\partial \mathcal{H}}{\partial z_j ^\ast},
 \]
% \end{equation}
where
\[
\frac{\partial}{\partial z^\ast}=
\frac{1}{2} \left( 
\frac{\partial}{\partial x}+i \frac{\partial}{\partial y} 
\right).
\]

For the leapfrogging problem, it is convenient to label the locations of the four vortices with the notation  $z_1^+, z_1^-, z_2^+,z_2^-$, which are assigned vorticities $\Gamma_{1,2}^+=1$ and $\Gamma_{1,2}^-=-1$~\cite{ArefTophoj}. The transformation
\begin{small}
\begin{equation}
    \begin{aligned}
\zeta&=\frac{1}{2}\left( z_1^+ +z_2^+-z_1^--z_2^-\right), &
\hat{\zeta}&=\frac{1}{2}\left( z_1^+ +z_2^+ + z_1^- + z_2^-\right),\\
\mathcal{Z} &=\frac{1}{2}\left( z_1^+ -z_2^+ + z_1^- - z_2^-\right), &
\mathcal{W} &=\frac{1}{2}\left( z_1^+ -z_2^+ - z_1^- + z_2^-\right),
\end{aligned}
\label{COV4to2}
\end{equation}
\end{small}
is canonical, i.e., it preserves the Hamiltonian form of the equations, as can be confirmed by computing the Poisson brackets of the new coordinates~\cite{Jose}. It is useful to reduce the number of degrees-of-freedom from four to two~\cite{ArefEckhardt}. In these variables $\mathcal{Z}$ is the vector connecting the centers of separations $d_1=z_1^+-z_1^-$ and $d_2=z_2^+-z_2^-$, whereas $\mathcal{W}=\frac{1}{2}(d_1-d_2)$ is half the difference between the two separations.
Further $\zeta=\frac{1}{2}(d_1+d_2)$ is one half the conserved linear impulse of the system and its conjugate $\hat{\zeta}$ is twice the centroid.

Following this transformation~\eqref{COV4to2}, the Hamiltonian 
%\begin{equation}\label{HamZW}
\[
\mathcal{H}= -\frac{1}{2\pi}
    \log{\abs{\frac{1}{\zeta^2-\mathcal{Z}^2}-\frac{1}{\zeta^2-\mathcal{W}^2}}}
\]
%\end{equation}
is cyclic in the variable $\hat{\zeta}$, which implies that $\zeta$ is conserved, i.e. $\zeta(t)=\zeta(0)=\zeta_0$.

By making appropriate scalings of both the independent and dependent variables (in the generic case $\zeta_0 \neq 0$), we arrive at the two degree-of-freedom Hamiltonian
\begin{align}\label{Hscaled}
\mathcal{H}(\mathcal{Z},\mathcal{W})=-\frac{1}{2} 
\log{\abs{\frac{1}{1+\mathcal{Z}^2}-\frac{1}{1+\mathcal{W}^2}}
}.
\end{align}
The evolution equations for the (complex-valued) coordinates $\mathcal{Z}$, $\mathcal{W}$ and the centroid $\hat{\zeta}$ are given by
%\begin{subequations}
\begin{align*} 
\overline{\frac{d \mathcal{Z}}{d t} }&=i \mathcal{W} \left( \frac{1}{\mathcal{Z}^2-\mathcal{W}^2}+\frac{1}{1+\mathcal{W}^2} \right),\\
\overline{\frac{d \mathcal{W}}{d t} }&=i \mathcal{Z} \left( \frac{1}{\mathcal{W}^2-\mathcal{Z}^2}+\frac{1}{1+\mathcal{Z}^2} \right),\\
\overline{\frac{d \hat{\zeta}}{d t} }&=\frac{1}{1+\mathcal{Z}^2}+\frac{1}{1+\mathcal{W}^2}.
\end{align*}
%\end{subequations}

Toph{\o}j notes that these are the canonical equations of motion not of Hamiltonian~\eqref{Hscaled} but of its extension to the \emph{complex-valued} Hamiltonian
\begin{align}\label{Hcomplex}
\mathscr{H}(\mathcal{Z},\mathcal{W})=
-\frac{1}{2} \log{\left(\frac{1}{1+\mathcal{Z}^2}-\frac{1}{1+\mathcal{W}^2} \right)},
\end{align}
which has equations of motion
\begin{align*}%\label{ComplexEofM}
\overline{\frac{d \mathcal{Z}}{dt}}= i\frac{\partial     \mathscr{H} }{\partial \mathcal{W} }, \quad \overline{\frac{d \mathcal{W}}{dt}}= i\frac{\partial     \mathscr{H} }{\partial \mathcal{Z}}.
\end{align*}
The topic of complex-valued Hamiltonians is not widely known, so we provide a reference~\cite{kaushal2000}.

\section{The leapfrogging solution and its linearization}\label{sec:leap_and_linear}

\subsection{Leapfrogging solutions}\label{sec:leapfrog_orbits}
At this point we find it preferable to again write the system in terms of real-valued coordinates and introduce the notation $\mathcal{Z}=X+i P$ and $\mathcal{W}=Q+i Y$. In these coordinates,  $(X,Q)$ is conjugate to $(Y,P)$. The subspace $P=Q=0$ is invariant under the motion, and corresponds exactly to the family of leapfrogging motions. On this invariant plane, $\mathcal{Z}=X$ and $\mathcal{W}=iY$ and the Hamiltonian~\eqref{Hcomplex} assumes only real values. The coordinates $(X,Y)$ evolve under the one degree-of-freedom Hamiltonian system with Hamiltonian,
\begin{align}
\label{1DoFHamiltonian}
H(X,Y)=\mathcal{H}(X,iY)=-\frac{1}{2} \log \left( \frac{1}{1-Y^2}-\frac{1}{1+X^2}   \right).
\end{align}

To simplify the mathematical analysis and allow the use of standard perturbation techniques, we make the following elementary observation. Given a Hamiltonian system with Hamiltonian $H(q,p)$, consider the modified system with Hamiltonian $\tilde{H}(q,p)= f\circ H(q,p)$, where $f\in C^1$ and is monotonic. Then the two systems have the same trajectories and equivalent dynamics up to a reparameterization of time by a factor of $f'(H)$. 

We apply this observation to the Hamiltonian~\eqref{1DoFHamiltonian}, which we note is singular at $(X,Y)=(0,0)$. This is the limit $\alpha$ approaches one, where the like-signed vortices coalesce into a single vortex with vorticity two. This causes the frequency of nearby oscillations to diverge to infinity. In order to desingularize the dynamics in this neighborhood, we redefine the Hamiltonian~\eqref{Hscaled} using  
\[
\tilde{\mathcal{H}}=f(\mathcal{H})=\frac{1}{2} e^ {-2\mathcal{H}},
\]
yielding the non-singular Hamiltonian in the invariant plane
\begin{equation}
\tilde{H}(X,Y) 
= \frac{1}{2} e^ {-2 H(X,Y)}
= \frac{1}{2} \left( \frac{1}{1-Y^2}-\frac{1}{1+X^2}   \right)
\label{Hnolog}
\end{equation}
and the new time scale
\begin{equation}
\tilde{t}=\frac{1}{f'(H)} t=-e^{2H}t.
\label{time_scale}
\end{equation}
The Hamiltonian in complex coordinates is regularized in the same manner,
\begin{equation}
  \tilde{\mathcal{H}}(\mathcal{Z},\mathcal{W})=f\left(\mathcal{H}\left(\mathcal{Z},\mathcal{W}\right)\right)=\frac{1}{2}\abs{\frac{1}{1+\mathcal{Z}^2}-\frac{1}{1+\mathcal{W}^2} }.
   \label{HWZnolog}
\end{equation}
For ease of notation, we will drop the tildes for the remainder of the paper. We also break with prior convention and use the value $h$ of the Hamiltonian $H$ in~\eqref{Hnolog} to parameterize the family of solutions, rather than using the ratio of the breadths of the vortex pairs, $\alpha$, as was done in previous work~\cite{Love,ArefTophoj,Acheson2000}.

With respect to energy level, $h$, leapfrogging motions occur for  $0< h<h_s=\frac{1}{2}$ and the leapfrogging motion has been found numerically to be stable for $0< h <h_{\rm c}=\frac{1}{8}$. The two parameters are related by $h= \frac{{(1-\alpha)}^2} {8 \alpha}$.

\begin{figure*}
  \centering
    \includegraphics[width=1\textwidth]{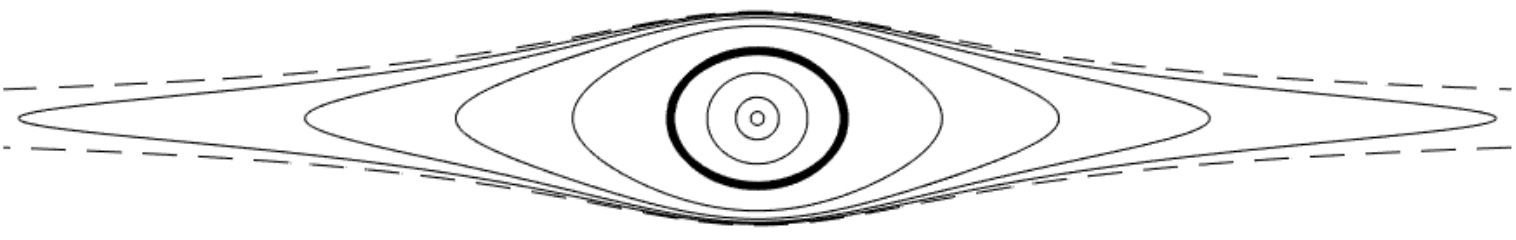}
    \caption{Level sets of the one-degree-of-freedom Hamiltonian~\eqref{Hnolog} in the $X-Y$ plane, including the critical energy level $H=\hc$ (bold) and the separatrix at $H=h_{\rm s}=\frac{1}{2}$ (dashed). Unbounded orbits not shown. The center at the origin corresponds $h=0$ in~\eqref{Hnolog} and to the limiting physical state in which the pairs of like-vorticity are at an infinitesimal distance and rotate with a divergent frequency as described by the original Hamiltonian. Stable orbits foliate the area between this point and the critical energy level.}
\label{LeapPPlane}
\end{figure*}

The Hamiltonian~\eqref{Hnolog} yields evolution equations
\begin{equation}
\label{EofM}
\begin{aligned}
\frac{dX}{dt}&=+\frac{\partial H}{\partial Y}=+\frac{Y}{{(1-Y^2)}^2},\\ 
\frac{dY}{dt}&=-\frac{\partial H}{\partial X}= -\frac{X}{{(1+X^2)}^2},
\end{aligned}
\end{equation}
whose phase plane is shown in Fig.~\ref{LeapPPlane}. In~\cite{grobli}, Gr\"{o}bli integrated the equations of motion~\eqref{EofM} to find an implicit formula for $X_h(t)$. In our notation, this is given by  
\begin{equation}
\label{implicit}
\begin{aligned}
 t(X)&=\frac{1}{2 h^2 \sqrt{1-4 h^2}}F\left(\sin ^{-1} \theta \big| k\right)-E\left(\sin ^{-1} \theta\big|k\right)\\
 &-\frac{1+2 h}{2 h \sqrt{(1-2 h) \left(2 h \left(X^2+1\right)+1\right)}},
\end{aligned}
\end{equation}
where $\theta=X\sqrt{\frac{2h-1}{2 h}}$, $k^2=\frac{4h^2}{4h^2-1}$, and $F$ and $E$ are incomplete elliptic integral of the first  and second kind respectively. To study the stability of these trajectories as solutions to~\eqref{HWZnolog}, it would be useful to  write them in an explicit closed form. Unfortunately,~\eqref{implicit} does not seem to be invertible to yield an explicit formula for $X(t)$. Nonetheless, in Sec.~\ref{sec:ExplicitFloquet} we reformulate the problem in order provides an explicit formulation of the stability problem without  having to invert this formula.

\subsection{Floquet theory and the linearized perturbation equations}\label{sec:Floquet_Pert}

To analyze the linear stability of the periodic orbit  $\gamma_h(t)=(X_h(t),Y_h(t),0,0)$, we perturb the evolution equations corresponding to Hamiltonian~\eqref{HWZnolog} about the leapfrogging solution $(X,Y,Q,P)=(X(t),Y(t),0,0)$. We introduce perturbation coordinates 
\begin{align*}
\mathcal{Z}(t)&= X(t)+\epsilon[ \xi_+(t)+i \eta_+(t) ],\\
\mathcal{W}(t)&= i Y(t)+\epsilon[ \xi_-(t)+i \eta_-(t) ],
\end{align*}
and expand the ODE system, keeping terms of linear order in $\epsilon$. The resulting equations decouple into two $2 \times 2$ systems,
\begin{subequations}
\label{PerturbationEquations}
\begin{align}
\label{eqA}
\frac{d}{dt}\,\, \big[\xi_+, \eta_-\big]^{\Tee}
&=A(X,Y)\,\,\,\,\big[\xi_+, \eta_-\big]^{\Tee}\quad\text{and}\\ 
\label{eqB}
\frac{d}{dt}\,\, \big[\xi_- , \eta_+  \big]^{\Tee}&=
A^{\Tee}(X,Y)\, \big[\xi_- ,\eta_+  \big]^{\Tee}.
\end{align}
\end{subequations}
where $A(X,Y)$ is given by
\[
A=\begin{pmatrix} 
 \frac{ XY}{(X^2+Y^2)(1+X^2)(1-Y^2)} & -\frac{3Y^4+X^2Y^2+X^2-Y^2}{2(X^2+Y^2){(1-Y^2)}^3} \\ 
 -\frac{3X^4+X^2Y^2-Y^2+X^2}{2(X^2+Y^2){(1+X^2)}^3} & -\frac{XY}{(X^2+Y^2)(1+X^2)(1-Y^2)} 
 \end{pmatrix}.
\]

Because these two systems depend only on quadratic terms in $(X,Y)$, the coefficient matrices have period $\frac{1}{2}T_{\rm leapfrog}$. Each is a linear Hamiltonian system since the matrix $A(t)$ on the right-hand side can be written as $A=JH$ where $J=\left(\begin{smallmatrix} 0 & 1 \\ -1 & 0 \end{smallmatrix} \right)$ and $H$ is symmetric.

In order to analyze these equations, we need to understand the behavior of solutions to the linear system with time-periodic coefficients, dependent on a parameter $\alpha$,
\begin{equation}
\label{floquet}
\dot{X}=A(t;\alpha) X, \quad A(t)=A(t+T;\alpha),
\end{equation} 
which is known as a Floquet problem~\cite{floquet1883equations,Meiss,Yakubovich}. To understand the behavior of solutions of equations of the form~\eqref{floquet}, we must review some basic facts from Floquet theory. Define the fundamental solution operator $\Phi(t)$ as the matrix-valued solution to~\eqref{floquet} with $\Phi (0)=I$. The \emph{monodromy matrix} is defined as the solution operator evaluated at one period  $M=\Phi(T)$. The eigenvalues, $\lambda$, of $M$ are called the Floquet multipliers. If any multiplier $\lambda$ satisfies $\abs{\lambda}>1$, then the solutions of the system of equations~\eqref{floquet} include an exponentially growing solution and the system is considered unstable.

If $A(t)$ is a $2\times2$ Hamiltonian matrix, the Floquet multipliers comes in pairs  $\lambda_1(\alpha)$ and $\lambda_2(\alpha)$ such that $\lambda_1\lambda_2=1$. If $\lambda_{1,2}$ have nonzero imaginary part, then the two multipliers must lie on the unit circle and be conjugate. If $\lambda_{1,2}$ are real and   $|\lambda_1|\neq 1$, then one multiplier lies inside the unit circle and the other lies outside the unit circle, and the system is unstable. On the boundary between stability and instability, the two eigenvalues must both lie on the unit circle and be real-valued, i.e., they must satisfy $\lambda_1=\lambda_2=\pm 1$.

The Floquet multipliers depend continuously on the parameter $\alpha$. Therefore, bifurcations, i.e., changes in stability, can only occur with $\lambda_1=\lambda_2=\pm 1$~\cite{Yakubovich}. The existence of a multiplier $\lambda=1$ (respectively $\lambda=-1$) corresponds to the existence of a periodic orbit with period $T$ (respectively, an anti-periodic orbit of half-period $T$). The stability or instability is easily determined by examining $\tr(M)=\lambda_1+\lambda_2$,  with stability in the case $\abs{\tr(M)}<2$ and instability when $\abs{\tr(M)}>2$. At the bifurcation values,  $\tr{M}=2$  and  $\tr{M}=-2$, the system~\eqref{floquet} has a periodic orbit or an anti-periodic orbit, respectively.

 We now return to the linearized perturbation equations of the leapfrogging orbit~\eqref{PerturbationEquations}. The coordinates $(\xi_+,\eta_-)$ describe perturbations within the family of periodic orbits. As such, the monodromy matrix for equation~\eqref{eqA} has eigenvalues $\lambda_{1,2}\equiv1$ which can lead to at most linear-in-time divergence of trajectories; see~\cite{ArefTophoj}. The question of stability is therefore determined entirely by the second system~\eqref{eqB}. Let $Z=(\xi_-, \eta_+)$, then~\eqref{eqB} can be written as 
\begin{equation}
\label{floquetOld}
\frac{d Z(t) }{dt}=A(X_h(t),Y_h(t)) Z(t),
\end{equation}
where 
\begin{equation*}
A(t)=A \left(t+\frac{1}{2}T_{\rm leapfrog}(h)\right),
\end{equation*}
and the period of the leapfrogging motion, $T_{\rm leapfrog}$, can be found from~\eqref{implicit} and is given by
\begin{equation*}%\label{Period}
T_{\rm leapfrog}(h)=\frac{8 h^2}{1-h^2} \left(h^2
  E\left(\frac{1}{h}\right)+\left(1-
  h^2\right)
  K\left(\frac{1}{h}\right)\right),
\end{equation*}
where $E$ and $K$ are complete elliptic integrals of the first and second kind respectively.

\section{Explicit form of the Floquet problem}
\label{sec:ExplicitFloquet}
\subsection{Reformulation in terms of the canonical polar angle}

The coordinates $X_h$ and $Y_h$ can not be solved in closed form. This is not a problem when finding the Floquet multipliers numerically, but it will be analytically useful to have an explicit form of the Floquet problem. To this end, we change the independent variable in a manner inspired by the proof that bounded solutions to the gravitational two-body problem are ellipses.
Consider the canonical polar coordinates~\cite{MeyerHall},
 \begin{equation}
\label{canonical_polar}
X=\sqrt{2J} \cos \theta,\qquad  
Y=\sqrt{2J} \sin \theta.
\end{equation}
This transformation preserves the Hamiltonian structure of the equations of motion, i.e.
\[
\frac{d\theta}{dt}=\frac{\partial H}{\partial J} 
\quad \textrm{and} \quad 
\frac{dJ}{dt}=-\frac{\partial H}{\partial \theta}.
\]
We rewrite~\eqref{floquetOld} as a Floquet problem with the polar angle $\theta$ as an independent variable. With respect to the variables $\theta$ and $J$, the  Hamiltonian~\eqref{Hnolog} can be rewritten as
\begin{equation*}%\label{1DoFHamiltonianPolar}
H(J,\theta)=\frac{2 J}{2-J^2-4 J \cos{2 \theta} +J^2 \cos{4 \theta} }.
\end{equation*}
At a given energy level $H=h$, we can solve for $J$,
\begin{equation}\label{jay}
J_{\pm}=\frac{1+2 h \cos{2 \theta} \pm\sqrt{1+4 h^2+4 h \cos{2 \theta} }}{h (-1+\cos{4 \theta}
   )}.
   \end{equation}
Of these two roots, only $J_{-}$ is both positive and free from singularities. Thus from here on, we set  $J=J_{-}(h,\theta)$. Since~\eqref{canonical_polar} is a canonical transformation, it preserves Hamilton's equations of motion. Therefore, $\theta$ evolves as
\begin{equation}
\label{dotTheta}
    \begin{aligned}
    \frac{d \theta}{dt}&=\frac{\partial H}{\partial J}=\frac{1}{2} \Big(1+4 h^2+4 h \cos{2 \theta}\\&+ (1+2h \cos{2 \theta})\sqrt{1+4 h^2+4 h \cos{2 \theta}}\Big).
    \end{aligned}
\end{equation}
where we have used~\eqref{jay} to write~\eqref{dotTheta} in terms of $h$ and $\theta$. 

In these variables, the Floquet matrix in~\eqref{floquetOld}, $A(J,\theta)$, is given by
\begin{small}
\begin{equation}
\label{AJ}
A=
\begin{pmatrix}
 \frac{- \sin{2 \theta}}{(-1+J+J \cos{2 \theta} )(-1-J+J \cos{2 \theta} )} &
  \frac{(2+6J) \cos{2 \theta}-J (5+\cos{4 \theta})}{{2(-1-J+J \cos{2 \theta})}^3} \\
 \frac{(2-6 J) \cos{2 \theta}-J (5+\cos{4 \theta})}{{2(-1+J+J \cos{2 \theta} )}^3} & 
 \frac{ \sin{2 \theta}}{(-1+J+J \cos{2 \theta} )(-1-J+J \cos{2 \theta} )} 
\end{pmatrix}.
\end{equation}
\end{small}

Using~\eqref{jay}, $J$ can be eliminated from $A(J,\theta)$ and~\eqref{AJ} can be written as a function $A_h(\theta)$,  depending on the parameter $h$ alone. Since 
\begin{equation}
\frac{d Z(\theta) }{d\theta}\frac{d\theta}{dt}=A_h(\theta) Z(\theta),
\end{equation}
equation~\eqref{dotTheta} can be used to write this as 
\begin{equation}
\label{floquet_theta}
\frac{d Z(\theta) }{d\theta}=\tilde{A}_h(\theta) Z(\theta) \quad \textrm{where} \quad \tilde{A}_h(\theta)={\left( \frac{d \theta}{dt}\right)}^{-1} A_h(\theta).
\end{equation}
In what follows, we drop the tilde from this notation.

In particular, at the apparent bifurcation value $h=1/8$, the coefficient matrix is given by
\begin{widetext}
\begin{small}
\begin{equation} \label{A_one_eighth}
    A_{h=\frac{1}{8}} (\theta)=
    \frac{1}{4\sqrt{17+8 \cos{2 \theta} }}
\begin{pmatrix}
 - \sin{2 \theta}  &
 \frac{7+12 \cos{2 \theta} -4 \cos{4 \theta}-3\sqrt{17+8 \cos{2 \theta} } }{2-2\cos{2 \theta}} \\
 \frac{3-4 \cos{2 \theta} -4\cos{4 \theta}-\sqrt{17+8 \cos2 \theta }} {2+2\cos 2\theta}&
  \sin{2 \theta}  \\
\end{pmatrix}.
\end{equation}
\end{small}
\end{widetext}
An additional benefit is that in this approach, the period is independent of $h$ since  $A_h(\theta)=A_h(\theta+\pi)$.

\subsection{Numerical solution of the Floquet problem}\label{sec:numerics}

\begin{figure*}[htbp] %  figure* placement: here, top, bottom, or page
   \centering
   \includegraphics[width=1\textwidth]{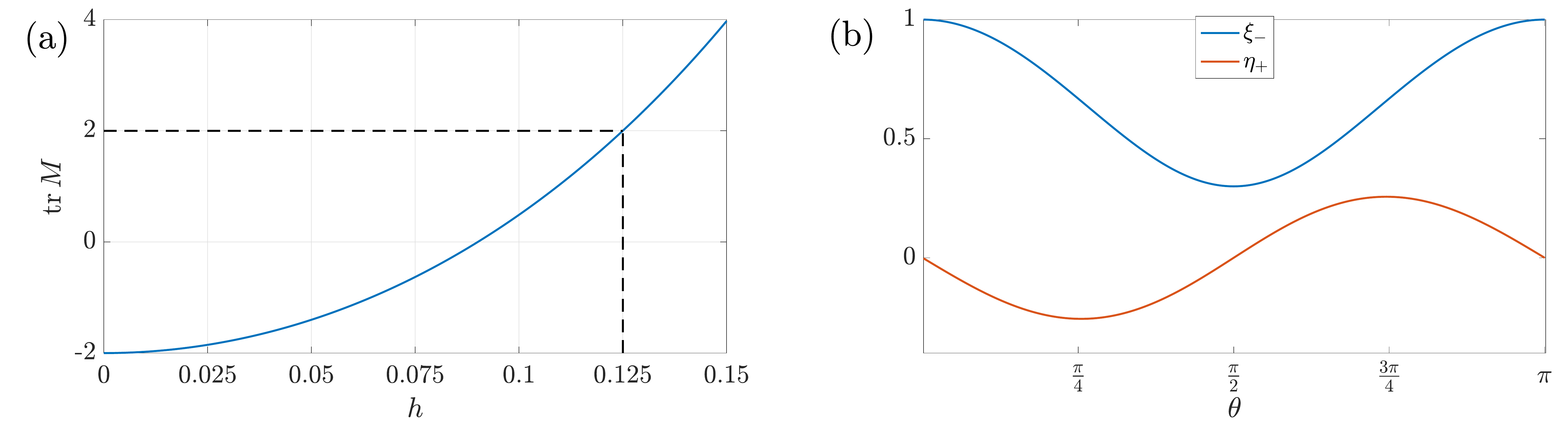} 
   \caption{(a) The trace of the monodromy matrix as a function of the energy $h$.  (b) The periodic orbit at $h=\frac{1}{8}$.}
  \label{fig:trace_periodic}
   \end{figure*}

Using this explicit construction, we give two numerical  checks for the critical value of $\hc=\frac{1}{8}$. Let $M_h$ be the monodromy matrix of the system~\eqref{floquet_theta}, and define the function $f(h)=\tr{M_h}-2$. 
We used   MATLAB's built in rootfinder, \texttt{fzero} along with the ODE Solver \texttt{ode45} with a relative tolerance of $10^{-13}$, an absolute tolerance of $10^{-15}$ to solve the equation $f(\hc) =0$. Using  an initial value of $h=0.1$, the solver returned the numerical solution $\hc=0.125$ to within machine error. Note that constructing $f(h)$ requires the numerical solution of the Floquet problem.  See Fig.~\ref{fig:trace_periodic}(a).

Another test, which is more relevant for the approach used in Sec.~\ref{sec:HillHBM}, is to check that the solution to~\eqref{A_one_eighth} has a periodic solution with an initial value of $Z(\theta)={(1, 0)}^{\Tee}$. In this formulation only a single system of two ODEs must be integrated. Using arbitrary precision arithmetic and a 30th order Taylor method using the Julia package \texttt{TaylorIntegration.jl}~
\cite{perez_hernandez_jorge_a_2019_2562353}, we find that the numerical solution satisfies   $||Z(\pi)-Z(0)||_2 <10^{-120}$. This is consistent with the hypothesis that $Z$ has a periodic solution of period $\pi$ and that $\hc$ is truly  rational up to the accuracy of the simulation. See Fig.~\ref{fig:trace_periodic}(b).

\subsection{Expansion in $h$}
The method of harmonic balance used in  Sec.~\ref{sec:HillHBM} requires that the Floquet matrix $A_h(\theta)$, with explicit form~\eqref{floquet_theta}, be written as a Fourier series. To accomplish this we expand $A_h$ in a Maclaurin series in $h$ and find at each order in $h$ a finite Fourier expansion. Letting
%\begin{equation}\label{A_h_expansion}
\[
A_h(\theta)=\sum_{k=0}^{\infty} h^k A_k(\theta), 
\]
%\end{equation}
the first few terms are given by
\begin{small}
\begin{subequations}
\begin{align*}
A_0(\theta)&=
\begin{pmatrix}
 -\sin 2\theta & -\cos 2\theta \\
 -\cos 2\theta & \sin 2\theta \\
\end{pmatrix}, \\
A_1(\theta)&=\begin{pmatrix}
 \sin{4 \theta} & 3+ \cos{4 \theta}\\
3+  \cos{4 \theta} & - \sin{4 \theta}\\
\end{pmatrix}, \\
A_2(\theta)&=\frac{1}{2} \begin{pmatrix}
 \sin{2 \theta} -3 \sin{6 \theta}  & 
 -12-9 \cos{2 \theta} -3\cos 6\theta  \\
 12+9 \cos{2 \theta} -3\cos{6 \theta} & 
 -\sin{2 \theta} +3 \sin{6 \theta}  \\
\end{pmatrix}.
\end{align*}
\end{subequations}
\end{small}

To perform a perturbation expansion, it is preferable that the leading-order term has constant-valued coefficients. The system can be put in such a form by a $\theta$-dependent change of variables known as a Lyapunov transformation, which we construct. First note that the matrix 
\begin{equation}\label{lyapunov}
B(\theta)=
\begin{pmatrix}
 \cos \theta & -\sin \theta \\
 \sin \theta & \cos \theta \\
\end{pmatrix},
\end{equation}
is the fundamental solution matrix of the system
\[
\frac{d B}{d\theta}=A_0 B.
\]
 Under the canonical change of variables $W(\theta)=B(\theta) Z(\theta)$, 
the system~\eqref{floquetOld} becomes
\begin{align*}
\frac{d W} {d\theta}&= \frac{dB}{d\theta} Z+B \frac{dZ}{d\theta} \\
&= \frac{dB}{d\theta} B^{-1} W+ B  A Z \\
&= \left( \frac{dB}{d\theta} B^{-1} + B  A B^{-1} \right) W.
\end{align*}
Letting 
\[
C(\theta)=\frac{dB}{d\theta} B^{-1} + B A B^{-1},
\] 
then 
\begin{equation}
\label{mainC}
\frac{d W} {d\theta}=C_h(\theta)W,
\end{equation}
 where the first few terms in the series
\begin{equation}\label{C_h_expansion}
C_h(\theta)=C_0+\sum_{k=1}^{\infty} h^k C_k(\theta) 
\end{equation}
are 
\begin{equation*}
    \begin{aligned}
C_0(\theta)&=
\begin{pmatrix}
 0 & -2 \\
 0 & 0 \\
\end{pmatrix}, \\ 
C_1(\theta)&=\begin{pmatrix}
 -2 \sin{2 \theta} & 4 \cos{2 \theta}\\
 4 \cos{2 \theta}  & 2 \sin{2 \theta}\\
\end{pmatrix},\\ 
C_2(\theta)&=\begin{pmatrix}
 \sin{4 \theta}       &  -8-\cos{4 \theta}  \\
 4 -4\cos{4 \theta}  & -\sin{4 \theta}     \\
\end{pmatrix}, \\ 
C_3(\theta)&=\begin{pmatrix}
 -5 \sin{2 \theta} -\sin{6 \theta}  & 26 \cos{2 \theta}+6 \cos{6 \theta}  \\
 -6 \cos 2\theta +6 \cos 6\theta  & 5 \sin 2\theta+\sin{6 \theta}      \\
   \end{pmatrix},
\end{aligned}
\end{equation*}
including a leading-order term that is independent of $\theta$, as desired.

\section{Method of harmonic balance and the  Hill's determinant}\label{sec:HillHBM}
In this section, we apply the method of harmonic balance (MHB) to the $\pi$-periodic differential equation~\eqref{mainC}. As noted in Sec.~\ref{sec:Floquet_Pert}, at parameter values  where the system undergoes a bifurcation, there must exist either a periodic  orbit or an anti-periodic orbit. The idea behind this method is that if such an orbit exists, then it has a convergent Fourier series which can be found if an approximate solvability condition for its coefficients is satisfied. In this section, we provide a brief overview of the method. For a thorough classical overview, see~\cite{whittaker1996}.

\subsection{Hill's formula} 
% MHB is a classical method due to G. W. Hill.

In his 1886 account of the motion of the lunar perigee~\cite{hill1886}, Hill considered what has come to be known as Hill's equation
\begin{equation}
   \label{Hill_second}
    \ddot{x}=g_\alpha(t) x(t),\text{ where } g_\alpha(t+2 \pi)=g(t).
\end{equation}
This can be put in the standard Floquet form~\eqref{floquet} with coefficient matrix $A(t,\alpha)=\left(\begin{smallmatrix} 0 & 1 \\ g_\alpha(t) & 0\end{smallmatrix}\right)$.
Hill formally found a relationship between the trace of the required mondromy matrix $M_\alpha$ and the coefficients forming the Fourier series of $g_\alpha(t)$. Hill's result, in a modern notation, can be summarized as follows. If $g_\alpha$ has Fourier expansion, 
\begin{equation}
   \label{g_fourier_expansion}
    g_\alpha(t)=\sum_{k =-\infty}^\infty g_k(\alpha) e^{ikt}, \qquad g_k\in \mathbb{C},
\end{equation}
then the infinite matrix, $H_{\alpha}=\left( h_{mk} (\alpha) \right)$, with components
\begin{equation}
\label{h_jk_reg}
    h_{mk}(\alpha)=\frac{k^2 \delta_{m k}+g_{k-m}(\alpha)}{k^2+1}, \qquad m,k \in \mathbb{Z},
\end{equation}
where $\delta_{ij}$ is the Kronecker delta, has determinant
\begin{equation}
   \label{hills_formula}
   \abs{H_\alpha}=\frac{\tr(M_{\alpha})-2}{e^{2\pi} +e^{-2\pi}-2}.
\end{equation}
 Notice that if the system~\eqref{Hill_second} has a periodic orbit at parameter value $\alpha$, then $\tr(M_{\alpha})=2
 $ and $\abs{H}=0$.

In 1899, Poincare proved the convergence of Hill's formula and gave a rigorous definition of the determinant of the infinite matrix $H_\alpha$~\cite{Poincare1899}. Hill's infinite determinant can also be given a variational interpretation as the Hessian of the action functional evaluated at the critical value given by the periodic orbit. This can provide useful information regarding the stability of the periodic solution via the Morse index~\cite{treschev2009, bolotin2010}.

\subsection{Using Hill's determinant to detect bifurcations}
In the study of bifurcations, the vanishing of Hill's determinant has a natural interpretation: it is a solvability condition for the values of the parameter $\alpha$ at which there exists either a periodic  orbit or an anti-periodic orbit, which indicates that the system may undergo a change of stability. The most familiar example of an equation in Hill's form is Mathieu's equation, in which the coefficient takes the form $g=c+d\cos {2t}$. Consider the ansatz where the solution to~\eqref{Hill_second}, $x(t)$, is  periodic and has Fourier expansion
\begin{equation}
   \label{complex_ansatz}
        x(t)=\sum_{k=-\infty}^\infty x_m e^{imt}, \qquad x_j\in \mathbb{C}.
\end{equation}
We will seek  a solvability condition for the existence of a non-trivial solution, $x(t)$. Putting~\eqref{g_fourier_expansion} and~\eqref{complex_ansatz} into~\eqref{Hill_second} and collecting harmonics yields the formal series
\begin{equation}
\label{h_jk}
    \begin{aligned}
     - \ddot{x}(t)+g(t) x(t) &= \sum_{k=-\infty}^\infty \sum_{m=-\infty}^\infty \left( m^2+g_k(\alpha) e^{ikt}\right)x_m e^{i m t } \\
      &=\sum_{k,m=-\infty}^\infty \left( k^2 \delta_{mk}+g_{k-m}(\alpha) \right) x_m e^{imt}\\
      &\equiv \sum_{k,m=-\infty}^\infty h_{mk}(\alpha) x_m e^{imt}=0.
    \end{aligned}
\end{equation}
This defines a matrix of infinite order, $H(\alpha)=\left(h_{km}(\alpha)\right)$. 

Consider a sequence of finite-dimensional matrices $H_N^{\rm{trunc}}$ obtained by truncating this system at the $N$th harmonic, i.e., only including terms 
 $k$ and $m$ such that $-N\leq k, m \leq N$. The matrix $H_N^{\rm{trunc}}$ has dimension $(2 N+1) \times (2N+1)$. As $N\to\infty$, roots of the equation $\abs{H_N^{\rm{trunc}}(\alpha)}=0$ should converge to the roots of $\abs{H(\alpha)}$%
\footnote{Note that equations~\eqref{h_jk_reg} and~\eqref{h_jk} differ by a factor of $\frac{1}{1+k^2}$. This is a regularization factor to guarantee $h_{jj}=1$ and is necessary for~\eqref{hills_formula} to converge.}%
.

\subsection{Application of the method}
To apply the method of harmonic balance, we write  
the periodic solution to system~\eqref{mainC} as a Fourier series. The following two observations allow us to simplify the form of this series. First, we observe that the first component of numerical solution  $Z(\theta)$ computed in Sec.~\ref{sec:numerics} is an even function and the second component is an odd function. Second, because it has period $\pi$, its Fourier series contains only even harmonics. Noting the definition of $W(\theta)$ using~\eqref{lyapunov}, this implies that $W(\theta)$ has only odd harmonics in its Fourier expansion. These two facts imply that $W(\theta)={(\xi_h,\eta_h)}^\Tee$ has the following Fourier expansion%
\footnote{This expansion contains only one-fourth of the possible non-zeros terms, and was based on mere observation from numerical simulations. It would of course be possible to proceed with a more general Fourier ansatz. We have done this, and found that the computed Hill determinant factors into several terms. Of these terms, only the one corresponding to the above expansion ever vanishes, so that no generality has been lost.}%
\begin{subequations}
\begin{align}
\xi_h(\theta)&= \sum_{n=1}^{\infty} a_{n}(h)\cos{\left(2n-1\right) \theta  } \quad \text{and}\\
\eta_h(\theta)&=  \sum_{n=1}^{\infty} b_{n}(h) 
\sin {\left(2n-1\right)\theta}.
\end{align}
\label{real_ansatz}%
\end{subequations}
We found using a trigonometric basis here to be more natural than a complex exponential basis used in~\eqref{complex_ansatz}. Putting this ansatz into the Floquet system~\eqref{mainC} and collecting coefficients of the harmonics formally result in an infinite-dimensional matrix problem  $M(h) \mathbf{a}=0$ where $\mathbf{a}={[a_1,b_1, a_3, b_3,\ldots]}^\Tee$.

To follow the approach of Hill, we need to truncate the Fourier ansatz~\eqref{real_ansatz} to $1\le n\le N$. Simultaneously, we truncate the series~\eqref{C_h_expansion} to $0\le k\le N$,
%\begin{equation}
\[
    C^{(N)}_h(\theta)=C_0+\sum_{k=1}^{N} h^k C_k(\theta).
\]
%\end{equation}
We therefore consider the sequence of truncated linear systems $M^{(N)} \mathbf{a}_N=\mathbf{0}$, where 
\begin{equation*}
\mathbf{a}_N = 
{[a_1,b_1, a_2, b_2,\ldots, a_N, b_N]}^\Tee.
\end{equation*}
This has nontrivial solutions if and only if  $M^{(N)}(h)$ is singular, i.e., if $\abs{M^{(N)}(h)}=0$.  We have automated this procedure in Mathematica~\cite{Mathematica} and can compute the result at arbitrary truncation order. The first two such truncated systems are
 \begin{align*}
\abs{M^{(1)}}&= \begin{vmatrix}
      -1+h & 2+2 h \\
     -2 h & 1-h \\
    \end{vmatrix}=-1+6 h+3 h^2,\\
\abs{M^{(2)}}&= \begin{vmatrix}
 -1+h & 2+2 h+8 h^2 & -h-\frac{h^2}{2} & -2
   h-2 h^2 \\
 -2 h-4 h^2 & 1-h & -2 h+2 h^2 &
   -h+\frac{h^2}{2} \\
 h-\frac{h^2}{2} & -2 h-2 h^2 & -3 & 2+8
   h^2 \\
 -2 h+2 h^2 & h+\frac{h^2}{2} & -4 h^2 & 3
   \\
\end{vmatrix}\\
&=9-54 h-109 h^2-210 h^3-\frac{977
   h^4}{2}+\frac{1049 h^5}{2}\\
   &\phantom{=}+\frac{75h^6}{2}+1074 h^7+\frac{11233 h^8}{16}.
\end{align*}
The relevant root of $\abs{M^{(1)}}=0$ can be found in closed form, $\hc^{(1)} = 2/\sqrt{3}-1 \approx 0.1547$ , but the rest must be found numerically. 
We have calculated the roots for several values of $N$ and have tabulated them in Table~\ref{table:detroots}. By a least-squares fit we find that the error, $\abs{\hc^{(N)}-\frac{1}{8}}$, decays at a rate of about $4^{-N}$. 

\begin{table}
\begin{center}
 \begin{tabular}{||c | c  ||} 
 \hline
$N$ &  $\hc^{(N)}$  \\ [0.5ex] 
 \hline
$1$  &  $0.154700538379256$\\
$2$  &  $0.125362196172840$\\
$3$  &  $0.125302181592097$\\
$4$  &  $0.125039391697053$ \\
$\vdots$ & $\vdots$ \\
% 5  &  0.125013678063844\\
% 6  &  0.125002532983010\\
% 7  &  0.125000749452121\\
% 8  &  0.125000157555837\\
% 9  &  0.125000043690148\\
% 10  &  0.125000009756739\\
% 11  &  0.125000002617060\\
% 12  &  0.125000000604347\\
% 13  &  0.125000000158988\\
% 14  &  0.125000000037475\\
% 15  &  0.125000000009738\\
% 16  &  0.125000000002326\\
% 17  &  0.125000000000599\\
% 18  &  0.125000000000145\\
% 19  &  0.125000000000037\\
$20$  &  $0.125000000000009$\\
  \hline 
\end{tabular} 
\end{center}
\caption{Roots of the truncated Hill determinants.}
\label{table:detroots}
\end{table}

\section{Conclusions}\label{sec:conclusion}

The present paper represents our  attempt to explain the fortuitous bifurcation value. Toward that end, we have derived an explicit reformulation of the stability problem, equation~\eqref{floquet_theta}. This is achieved by a transformation used in solving the Kepler problem~\cite{Jose}. This formulation allows us to pose the stability problem with periodic coefficients that are given exactly, whereas previous studies considered linearizing about a numerical solution. This simplified problem allows us to show, numerically, that there is a solution that is periodic to within an error on the scale of $10^{-120}$  

We then expand  system in a Fourier-Taylor series, using the energy $h$ as a small parameter. We employ a classical technique from the study of lunar motion due to G. W. Hill, which uses the method of harmonic balance, to derive a sequence of algebraic criteria for the stability of the leapfrogging orbits. The roots of these polynomials form a sequence of approximations that appears to converge exponentially to $\hc$.

We had hoped that this analysis would provide insight into a mechanism illuminating the surprising algebraic  critical value,  perhaps in the form of an exact formula for the periodic orbit. We do not yet see how this is possible, given the coefficients in equation~\eqref{A_one_eighth}. However, this approach shows an application of how ideas from the gravitational $N$-body problem can be transferred to the $N$-vortex problem.

For example, several generalizations of the leapfrogging solution exist and may be amenable to the techniques discussed here. First, leapfrogging solutions exist for quartets consisting of two pairs with vorticities  $\Gamma_1^- = -\Gamma_1^+$ and $\Gamma_2^- = - \Gamma_2^+$. This reduces to the case  studied here when $\Gamma_1^+=\Gamma_2^+$. In the more general case the critical energy level should now depend on the ratio of the vorticities, $\lambda=\frac{\Gamma_1}{\Gamma_2}$.  Acheson reports that he has investigated this situation numerically through direct simulations~\cite{Acheson2000}. He makes a few observations about the behavior and suggests that it would be worthwhile to conduct a systematic analysis. We believe the semi-analytic method here is especially well suited for such as analysis as it will allow us to build the stability curves in $(h,\lambda)$ space.

% valid in a neighborhood of the walkabout and braiding solutions and given by a singular perturbation to a three-vortex system studied by Rott and by Aref~\cite{Rott,Aref1989}. We hope to use this system to shed additional light on the situation. More generally, braiding and walkabout solutions arise as intermediate states in the chaotic scattering of vortex dipoles, and we hope this reduction can be used to study that problem as well~\cite{Aref2008,Price}.

Another generalization is that leapfrogging solutions  exist for a system of $2N$ vortices with $N>2$, half with vorticity $+1$ and half with vorticity $-1$. As the leapfrogging of four vortices models the leapfrogging of two vortex rings, so the leapfrogging of $2N$ vortices models the leapfrogging of $N$ vortex rings, a problem that has been studied experimentally in superfluid helium. The latter system has been studied by Wacks et al.~\cite{Wacks:2014}. While they found the motion to be stable in their numerical simulations, reduction to an ODE system would allow the exploration of a larger volume of parameter space and the application of more theoretical tools. A third generalization is to consider a system of vortices confined to a sphere, in this case, the leapfrogging solution is symmetric about a great circle. P. Newton has simulated these solutions, but their stability has not been analyzed~\cite{newton2008}.

Finally, we remark that we have not addressed the question of nonlinear dynamics of linearly unstable leapfrogging orbits, for example the transition from leapfrogging to walkabout and braiding orbits and even disintegration and escape. This will be the topic of an upcoming paper.

\begin{acknowledgments}
We thank Panos Kevrekidis for introducing us to this problem in a conversation made possible by a 2015 workshop at Dalhousie University sponsored by AARMS and arranged by Theodore Kolokolnikov, and for many subsequent discussions. We thank Stefanella Boatto,
Alain Brizard,
Jared Bronski,
Kevin Mitchell,
Gareth Roberts, 
Vered Rom-Kedar,
Spencer Smith,
and
Cristina Stoica
for additional useful discussions.
\end{acknowledgments}
\bibliography{Brandon_Leap_Frog.bib}{}

\begin{thebibliography}{35}
\expandafter\ifx\csname natexlab\endcsname\relax\def\natexlab#1{#1}\fi
\expandafter\ifx\csname bibnamefont\endcsname\relax
  \def\bibnamefont#1{#1}\fi
\expandafter\ifx\csname bibfnamefont\endcsname\relax
  \def\bibfnamefont#1{#1}\fi
\expandafter\ifx\csname citenamefont\endcsname\relax
  \def\citenamefont#1{#1}\fi
\expandafter\ifx\csname url\endcsname\relax
  \def\url#1{\texttt{#1}}\fi
\expandafter\ifx\csname urlprefix\endcsname\relax\def\urlprefix{URL }\fi
\providecommand{\bibinfo}[2]{#2}
\providecommand{\eprint}[2][]{\url{#2}}

\bibitem[{\citenamefont{von Helmholtz}(1858)}]{Helmholtz}
\bibinfo{author}{\bibfnamefont{H.}~\bibnamefont{von Helmholtz}},
  \bibinfo{journal}{J. Reine Angew. Math} \textbf{\bibinfo{volume}{55}},
  \bibinfo{pages}{25} (\bibinfo{year}{1858}).

\bibitem[{\citenamefont{Newton}(2013)}]{newton2013n}
\bibinfo{author}{\bibfnamefont{P.~K.} \bibnamefont{Newton}},
  \emph{\bibinfo{title}{The $N$-Vortex Problem: Analytical Techniques}},
  Applied Mathematical Sciences (\bibinfo{publisher}{Springer New York},
  \bibinfo{year}{2013}), ISBN \bibinfo{isbn}{9781468492903}.

\bibitem[{\citenamefont{Aref}(2007)}]{Aref2007}
\bibinfo{author}{\bibfnamefont{H.}~\bibnamefont{Aref}},
  \bibinfo{journal}{Journal of Mathematical Physics}
  \textbf{\bibinfo{volume}{48}}, \bibinfo{pages}{065401}
  (\bibinfo{year}{2007}), ISSN \bibinfo{issn}{0022-2488}.

\bibitem[{\citenamefont{Aref}(2010)}]{Aref150}
\bibinfo{author}{\bibfnamefont{H.}~\bibnamefont{Aref}},
  \bibinfo{journal}{Theor. Comput. Fluid Dyn.} \textbf{\bibinfo{volume}{24}},
  \bibinfo{pages}{1} (\bibinfo{year}{2010}), ISSN \bibinfo{issn}{1432-2250}.

\bibitem[{\citenamefont{Kirchhoff}(1876)}]{kirchhoff1876}
\bibinfo{author}{\bibfnamefont{G.}~\bibnamefont{Kirchhoff}},
  \emph{\bibinfo{title}{Vorlesungen {\"u}ber mathematische Physik: Mechanik}},
  vol.~\bibinfo{volume}{1} of \emph{\bibinfo{series}{Vorlesungen {\"u}ber
  mathematische Physik}} (\bibinfo{publisher}{Teubner, Leipzig},
  \bibinfo{year}{1876}).

\bibitem[{\citenamefont{Aref et~al.}(1992)\citenamefont{Aref, Rott, and
  Thomann}}]{Aref1992}
\bibinfo{author}{\bibfnamefont{H.}~\bibnamefont{Aref}},
  \bibinfo{author}{\bibfnamefont{N.}~\bibnamefont{Rott}}, \bibnamefont{and}
  \bibinfo{author}{\bibfnamefont{H.}~\bibnamefont{Thomann}},
  \bibinfo{journal}{Annu. Rev. Fluid Mech.} \textbf{\bibinfo{volume}{24}},
  \bibinfo{pages}{1} (\bibinfo{year}{1992}), ISSN \bibinfo{issn}{0066-4189}.

\bibitem[{\citenamefont{Eckhardt and Aref}(1988)}]{ArefEckhardt}
\bibinfo{author}{\bibfnamefont{B.}~\bibnamefont{Eckhardt}} \bibnamefont{and}
  \bibinfo{author}{\bibfnamefont{H.}~\bibnamefont{Aref}},
  \bibinfo{journal}{Philos. Trans. R. Soc. London, Ser. A}
  (\bibinfo{year}{1988}).

\bibitem[{\citenamefont{Aref}(1989)}]{Aref1989}
\bibinfo{author}{\bibfnamefont{H.}~\bibnamefont{Aref}},
  \bibinfo{journal}{Journal of Applied Mathematics and Physics}
  (\bibinfo{year}{1989}).

\bibitem[{\citenamefont{Anderson et~al.}(1995)\citenamefont{Anderson, Ensher,
  Matthews, Wieman, and Cornell}}]{anderson1995}
\bibinfo{author}{\bibfnamefont{M.~H.} \bibnamefont{Anderson}},
  \bibinfo{author}{\bibfnamefont{J.~R.} \bibnamefont{Ensher}},
  \bibinfo{author}{\bibfnamefont{M.~R.} \bibnamefont{Matthews}},
  \bibinfo{author}{\bibfnamefont{C.~E.} \bibnamefont{Wieman}},
  \bibnamefont{and} \bibinfo{author}{\bibfnamefont{E.~A.}
  \bibnamefont{Cornell}}, \bibinfo{journal}{Science}
  \textbf{\bibinfo{volume}{269}}, \bibinfo{pages}{198} (\bibinfo{year}{1995}).

\bibitem[{\citenamefont{Matthews et~al.}(1999)\citenamefont{Matthews, Anderson,
  Haljan, Hall, Wieman, and Cornell}}]{Matthews1999}
\bibinfo{author}{\bibfnamefont{M.~R.} \bibnamefont{Matthews}},
  \bibinfo{author}{\bibfnamefont{B.~P.} \bibnamefont{Anderson}},
  \bibinfo{author}{\bibfnamefont{P.~C.} \bibnamefont{Haljan}},
  \bibinfo{author}{\bibfnamefont{D.~S.} \bibnamefont{Hall}},
  \bibinfo{author}{\bibfnamefont{C.~E.} \bibnamefont{Wieman}},
  \bibnamefont{and} \bibinfo{author}{\bibfnamefont{E.~A.}
  \bibnamefont{Cornell}}, \bibinfo{journal}{Phys. Rev. Lett.}
  \textbf{\bibinfo{volume}{83}}, \bibinfo{pages}{2498} (\bibinfo{year}{1999}).

\bibitem[{\citenamefont{Navarro et~al.}(2013)\citenamefont{Navarro,
  Carretero-Gonz{\'a}lez, Torres, Kevrekidis, Frantzeskakis, Ray, Altunta{\c
  s}, and Hall}}]{Navarro:2013hb}
\bibinfo{author}{\bibfnamefont{R.}~\bibnamefont{Navarro}},
  \bibinfo{author}{\bibfnamefont{R.}~\bibnamefont{Carretero-Gonz{\'a}lez}},
  \bibinfo{author}{\bibfnamefont{P.~J.} \bibnamefont{Torres}},
  \bibinfo{author}{\bibfnamefont{P.~G.} \bibnamefont{Kevrekidis}},
  \bibinfo{author}{\bibfnamefont{D.~J.} \bibnamefont{Frantzeskakis}},
  \bibinfo{author}{\bibfnamefont{M.~W.} \bibnamefont{Ray}},
  \bibinfo{author}{\bibfnamefont{E.}~\bibnamefont{Altunta{\c s}}},
  \bibnamefont{and} \bibinfo{author}{\bibfnamefont{D.~S.} \bibnamefont{Hall}},
  \bibinfo{journal}{Phys. Rev. Lett.} \textbf{\bibinfo{volume}{110}},
  \bibinfo{pages}{225301} (\bibinfo{year}{2013}).

\bibitem[{\citenamefont{Gr\"{o}bli}(1877)}]{grobli}
\bibinfo{author}{\bibfnamefont{W.}~\bibnamefont{Gr\"{o}bli}}, Ph.D. thesis,
  \bibinfo{school}{Georg-August-Universit\"{a}t G\"{o}ttingen}
  (\bibinfo{year}{1877}).

\bibitem[{\citenamefont{Love}(1893)}]{Love}
\bibinfo{author}{\bibfnamefont{A.~E.~H.} \bibnamefont{Love}},
  \bibinfo{journal}{Proc. London Math. Soc.} \textbf{\bibinfo{volume}{1}},
  \bibinfo{pages}{185} (\bibinfo{year}{1893}).

\bibitem[{\citenamefont{Borisov et~al.}(2005)\citenamefont{Borisov, Kilin, and
  Mamaev}}]{Borisov}
\bibinfo{author}{\bibfnamefont{A.~V.} \bibnamefont{Borisov}},
  \bibinfo{author}{\bibfnamefont{A.~A.} \bibnamefont{Kilin}}, \bibnamefont{and}
  \bibinfo{author}{\bibfnamefont{I.~S.} \bibnamefont{Mamaev}},
  \bibinfo{journal}{Discrete Contin. Dyn. Syst.} \textbf{\bibinfo{volume}{54}},
  \bibinfo{pages}{100} (\bibinfo{year}{2005}), ISSN \bibinfo{issn}{1078-0947}.

\bibitem[{\citenamefont{Acheson}(2000)}]{Acheson2000}
\bibinfo{author}{\bibfnamefont{D.~J.} \bibnamefont{Acheson}},
  \bibinfo{journal}{Eur. J. Phys.} \textbf{\bibinfo{volume}{21}},
  \bibinfo{pages}{269} (\bibinfo{year}{2000}), ISSN \bibinfo{issn}{0143-0807}.

\bibitem[{\citenamefont{Toph{\o}j and Aref}(2008)}]{Aref2008}
\bibinfo{author}{\bibfnamefont{L.}~\bibnamefont{Toph{\o}j}} \bibnamefont{and}
  \bibinfo{author}{\bibfnamefont{H.}~\bibnamefont{Aref}},
  \bibinfo{journal}{Phys. Fluids} \textbf{\bibinfo{volume}{20}},
  \bibinfo{pages}{093605} (\bibinfo{year}{2008}), ISSN
  \bibinfo{issn}{1070-6631}.

\bibitem[{\citenamefont{Toph{\o}j and Aref}(2013)}]{ArefTophoj}
\bibinfo{author}{\bibfnamefont{L.}~\bibnamefont{Toph{\o}j}} \bibnamefont{and}
  \bibinfo{author}{\bibfnamefont{H.}~\bibnamefont{Aref}},
  \bibinfo{journal}{Phys. Fluids} \textbf{\bibinfo{volume}{25}},
  \bibinfo{pages}{014107} (\bibinfo{year}{2013}), ISSN
  \bibinfo{issn}{1070-6631}.

\bibitem[{\citenamefont{Meiss}(2007)}]{Meiss}
\bibinfo{author}{\bibfnamefont{J.~D.} \bibnamefont{Meiss}},
  \emph{\bibinfo{title}{Differential Dynamical Systems}}
  (\bibinfo{publisher}{SIAM}, \bibinfo{address}{Philadelphia},
  \bibinfo{year}{2007}).

\bibitem[{\citenamefont{Markus and Yamabe}(1960)}]{markus1960global}
\bibinfo{author}{\bibfnamefont{L.}~\bibnamefont{Markus}} \bibnamefont{and}
  \bibinfo{author}{\bibfnamefont{H.}~\bibnamefont{Yamabe}},
  \bibinfo{journal}{Osaka J. Math.} \textbf{\bibinfo{volume}{12}},
  \bibinfo{pages}{305} (\bibinfo{year}{1960}).

\bibitem[{\citenamefont{Whitchurch et~al.}(2018)\citenamefont{Whitchurch,
  Kevrekidis, and Koukouloyannis}}]{Whitchurch}
\bibinfo{author}{\bibfnamefont{B.}~\bibnamefont{Whitchurch}},
  \bibinfo{author}{\bibfnamefont{P.~G.} \bibnamefont{Kevrekidis}},
  \bibnamefont{and}
  \bibinfo{author}{\bibfnamefont{V.}~\bibnamefont{Koukouloyannis}},
  \bibinfo{journal}{Phys. Rev. Fluids} \textbf{\bibinfo{volume}{3}},
  \bibinfo{pages}{014401} (\bibinfo{year}{2018}).

\bibitem[{\citenamefont{Rott}(1989)}]{Rott}
\bibinfo{author}{\bibfnamefont{N.}~\bibnamefont{Rott}}, \bibinfo{journal}{J.
  Appl. Math. Phys. (ZAMP)} \textbf{\bibinfo{volume}{40}}, \bibinfo{pages}{473}
  (\bibinfo{year}{1989}).

\bibitem[{\citenamefont{Jose and Saletan}(1998)}]{Jose}
\bibinfo{author}{\bibfnamefont{J.~V.} \bibnamefont{Jose}} \bibnamefont{and}
  \bibinfo{author}{\bibfnamefont{E.~J.} \bibnamefont{Saletan}},
  \emph{\bibinfo{title}{Classical Dynamics: A Contemporary Approach}}
  (\bibinfo{publisher}{Cambridge University Press}, \bibinfo{year}{1998}).

\bibitem[{\citenamefont{Kaushal and Korsch}(2000)}]{kaushal2000}
\bibinfo{author}{\bibfnamefont{R.}~\bibnamefont{Kaushal}} \bibnamefont{and}
  \bibinfo{author}{\bibfnamefont{H.}~\bibnamefont{Korsch}},
  \bibinfo{journal}{Phys. Lett. A} \textbf{\bibinfo{volume}{276}},
  \bibinfo{pages}{47} (\bibinfo{year}{2000}), ISSN \bibinfo{issn}{0375-9601}.

\bibitem[{\citenamefont{Floquet}(1883)}]{floquet1883equations}
\bibinfo{author}{\bibfnamefont{G.}~\bibnamefont{Floquet}}, in
  \emph{\bibinfo{booktitle}{Annales scientifiques de l'{\'E}cole normale
  sup{\'e}rieure}} (\bibinfo{year}{1883}), vol.~\bibinfo{volume}{12}, pp.
  \bibinfo{pages}{47--88}.

\bibitem[{\citenamefont{Yakubovich and Starzhinskii}(1975)}]{Yakubovich}
\bibinfo{author}{\bibfnamefont{V.}~\bibnamefont{Yakubovich}} \bibnamefont{and}
  \bibinfo{author}{\bibfnamefont{V.}~\bibnamefont{Starzhinskii}},
  \emph{\bibinfo{title}{Linear Differential Equations with Periodic
  Coefficients Vol. 1}} (\bibinfo{publisher}{John Wiley and Sons},
  \bibinfo{year}{1975}).

\bibitem[{\citenamefont{Meyer et~al.}(2008)\citenamefont{Meyer, Hall, and
  Offin}}]{MeyerHall}
\bibinfo{author}{\bibfnamefont{K.}~\bibnamefont{Meyer}},
  \bibinfo{author}{\bibfnamefont{G.}~\bibnamefont{Hall}}, \bibnamefont{and}
  \bibinfo{author}{\bibfnamefont{D.}~\bibnamefont{Offin}},
  \emph{\bibinfo{title}{Introduction to Hamiltonian Dynamical Systems and the
  N-body Problem}} (\bibinfo{publisher}{Springer}, \bibinfo{year}{2008}).

\bibitem[{\citenamefont{Pérez-Hernández and
  Benet}(2019)}]{perez_hernandez_jorge_a_2019_2562353}
\bibinfo{author}{\bibfnamefont{J.~A.} \bibnamefont{Pérez-Hernández}}
  \bibnamefont{and} \bibinfo{author}{\bibfnamefont{L.}~\bibnamefont{Benet}},
  \emph{\bibinfo{title}{{PerezHz/TaylorIntegration.jl: TaylorIntegration
  v0.4.1}}} (\bibinfo{year}{2019}),
  \urlprefix\url{https://doi.org/10.5281/zenodo.2562353}.

\bibitem[{\citenamefont{Whittaker and Watson}(1902)}]{whittaker1996}
\bibinfo{author}{\bibfnamefont{E.~T.} \bibnamefont{Whittaker}}
  \bibnamefont{and} \bibinfo{author}{\bibfnamefont{G.~N.}
  \bibnamefont{Watson}}, \emph{\bibinfo{title}{{A Course of Modern Analysis}}}
  (\bibinfo{publisher}{Cambridge University Press}, \bibinfo{year}{1902}).

\bibitem[{\citenamefont{Hill}(1886)}]{hill1886}
\bibinfo{author}{\bibfnamefont{G.~W.} \bibnamefont{Hill}},
  \bibinfo{journal}{Acta Math.} \textbf{\bibinfo{volume}{8}},
  \bibinfo{pages}{1} (\bibinfo{year}{1886}), ISSN \bibinfo{issn}{0001-5962}.

\bibitem[{\citenamefont{Poincar\'{e}}(1899)}]{Poincare1899}
\bibinfo{author}{\bibfnamefont{H.}~\bibnamefont{Poincar\'{e}}},
  \emph{\bibinfo{title}{Les m\'{e}thodes nouvelles de la m\'{e}canique
  c\'{e}leste. {T}ome {III}}} (\bibinfo{publisher}{Gauthier-Villars},
  \bibinfo{year}{1899}).

\bibitem[{\citenamefont{Treschev et~al.}(2009)\citenamefont{Treschev,
  Zubelevich, Treschev, and Zubelevich}}]{treschev2009}
\bibinfo{author}{\bibfnamefont{D.}~\bibnamefont{Treschev}},
  \bibinfo{author}{\bibfnamefont{O.}~\bibnamefont{Zubelevich}},
  \bibinfo{author}{\bibfnamefont{D.}~\bibnamefont{Treschev}}, \bibnamefont{and}
  \bibinfo{author}{\bibfnamefont{O.}~\bibnamefont{Zubelevich}},
  \emph{\bibinfo{title}{{Introduction to the Perturbation Theory of Hamiltonian
  Systems}}} (\bibinfo{publisher}{Springer-Verlag Berlin Heidelberg},
  \bibinfo{year}{2009}).

\bibitem[{\citenamefont{Bolotin and Treschev}(2010)}]{bolotin2010}
\bibinfo{author}{\bibfnamefont{S.}~\bibnamefont{Bolotin}} \bibnamefont{and}
  \bibinfo{author}{\bibfnamefont{D.}~\bibnamefont{Treschev}},
  \bibinfo{journal}{Russ. Math. Surv.} \textbf{\bibinfo{volume}{65}},
  \bibinfo{pages}{191} (\bibinfo{year}{2010}), ISSN \bibinfo{issn}{0036-0279}.

\bibitem[{\citenamefont{{Wolfram Research{,} Inc.}}(2019)}]{Mathematica}
\bibinfo{author}{\bibnamefont{{Wolfram Research{,} Inc.}}},
  \emph{\bibinfo{title}{Mathematica, {V}ersion 12.0}} (\bibinfo{year}{2019}),
  \bibinfo{note}{champaign, IL}.

\bibitem[{\citenamefont{Wacks et~al.}(2014)\citenamefont{Wacks, Baggaley, and
  Barenghi}}]{Wacks:2014}
\bibinfo{author}{\bibfnamefont{D.~H.} \bibnamefont{Wacks}},
  \bibinfo{author}{\bibfnamefont{A.~W.} \bibnamefont{Baggaley}},
  \bibnamefont{and} \bibinfo{author}{\bibfnamefont{C.~F.}
  \bibnamefont{Barenghi}}, \bibinfo{journal}{Phys. Fluids}
  \textbf{\bibinfo{volume}{26}}, \bibinfo{pages}{027102}
  (\bibinfo{year}{2014}).

\bibitem[{\citenamefont{Newton and Shokraneh}(2008)}]{newton2008}
\bibinfo{author}{\bibfnamefont{P.~K.} \bibnamefont{Newton}} \bibnamefont{and}
  \bibinfo{author}{\bibfnamefont{H.}~\bibnamefont{Shokraneh}},
  \bibinfo{journal}{Proc. R. Soc. London, Ser. A}
  \textbf{\bibinfo{volume}{464}}, \bibinfo{pages}{1525} (\bibinfo{year}{2008}),
  ISSN \bibinfo{issn}{1364-5021}.

\end{thebibliography}

\end{document}